\documentclass[fleqn,10pt]{wlscirep}
\usepackage[utf8]{inputenc}
\usepackage[T1]{fontenc}
\usepackage{booktabs}       
\usepackage{amsfonts}       
\usepackage{nicefrac}       
\usepackage{microtype}      
\usepackage{lipsum}
\usepackage{subfig}
\usepackage{graphicx}
\graphicspath{ {./figure/} }
\usepackage{stfloats}
\usepackage{amsmath}
\usepackage{amsmath,bm}
\title{An Irregular-shaped Ring-Pair Magnet Array with a Monotonic Field Gradient for 2D Head Imaging in Low-field Portable MRI}

\author[1]{Zhi Hua Ren}
\author[1]{Jia Gong}
\author[1,2,*]{Shao Ying Huang}

\affil[1]{Engineering Product Development, Singapore University of Technology and Design, 8 Somapah Road, Singapore 487372}
\affil[2]{Department of Surgery, National University of Singapore, IE Kent Ridge Road Singapore 119228}

\affil[*]{corresponding Author. Email: huangshaoying@sutd.edu.sg}

\begin{abstract}
We present a design and the optimization of an irregular-shaped ring-pair magnet array that generates a 1D monotonic field pattern for 2D head imaging in a low-field portable MRI system. The magnet rings are discretized into fan-shaped ring segments with varying outer diameters for the design and optimization. Besides, the inner radii of ring-pairs are tapered from outside in to provide the controlled field inhomogeneity. Genetic algorithm (GA) was used, and a current model for a fan-shaped ring segment was derived to have a fast forward calculation in the optimization. A monotonic field pattern is successfully obtained along the $x-$direction in a cylindrical field of view (FoV), with a relatively strong magnetic field (132.98\,mT) and the homogeneity of 151840\,ppm. The proposed array was further evaluated by applying its field as a spatial encoding magnetic field (SEM) for imaging numerically. Due to the field monotonicity, the reconstructed image by applying the fields of the proposed array shows clearer features (a higher structural similarity index) with a reduced error rate compared to that using a sparse dipolar Halbach array. The proposed magnet array is a promising alternative to supply SEM for imaging in a permanent-magnet-based low-field portable MRI system. 
\end{abstract}
\begin{document}

\flushbottom
\maketitle
%
%
\thispagestyle{empty}

\noindent 

\section{Introduction}
With no power consumption and low cost, permanent magnet arrays are always attractive to be used as a source of the static main magnetic field ($\bm{B}_0$ field) for a portable MRI scanner\,\cite{cooley2015,ren2017,shaoying2018}. However, when a traditional imaging approach is taken with Fourier transformation and linear gradient fields, homogeneous $\bm{B_0}$ fields are required in an MRI scanner, resulting in a bulky magnet array\,\cite{simens0p35T}, or if the magnet array is scaled down to a portable size, the imaging volume becomes too small to image a human organ\,\cite{huson2002_cshape_tabletop}. 
Recently, nonlinear gradient fields are proposed to be used as spatial encoding magnetic fields (SEMs) for MRI image reconstructions to overcome physiological limitations of the conventional spatial linear gradient setup, e.g. to reduce peripheral nerve stimulation. Some examples are the parallel imaging technique using localized gradients (PatLoc) imaging$\,$\cite{hennig2008,schultz2011radial}, and  O-space imaging$\,$\cite{stockmann2010space,stockmann2013vivo}.
In these approaches, as static field patterns without spatial linearity can be used to encode MRI signals for imaging, the requirements for the homogeneity of the static field can be relaxed. When a permanent magnet array is used to supply the $\bm{B_0}$ field, it allows an array with a reduced size, reduced weight to have a bigger imaging volume which may accommodate a part of human body, such as a head. It offers opportunities of constructing a truly portable low-cost MRI scanner. 

In\,\cite{cooley2015}, a sparse dipolar Halbach array\,\cite{halbach1979,halbach1980} (also named ``NMR Mandhala"\,\cite{Blumler2002_NMR_Mandhalas}) was used to supply a SEM (the main field plus the gradient field) with a quadrupolar field pattern in the transverse direction for head imaging with a portability.  
The SEM from a sparse Halbach array is curvilinear and nonbijective, so the sensitivity encoding (SENSE) using multiple receiver coils was applied to resolve the spatial ambiguity$\,$\cite{pruessmann1999sense,schultz2010}. Furthermore, the magnet array was rotated to obtain a variation of $\bf{B_0}$ field relative to the sample, so as to provide additional spatial encoding information to localize the MR signal$\,$\cite{cooley2015}. Although the ambiguity is mitigated by using multiple receive coils and by rotating the magnet, the central region of the quadrupolar SEMs is relatively homogeneous, which results in substantial blurring in that region of the imaging volume. Moreover, in the region with low or zero gradients, the image quality can greatly be degraded$\,$\cite{hennig2008,schultz2010reconstruction,cooley2015}. In\,\cite{ren2017}, it was reported that a shorter sparse Halbach array was implemented for MR imaging. Similar problems of imaging due to the nonbijective encoding $B_0$-fields were encountered. In such an MRI system where a permanent magnet array is used, monotonic SEMs with a relatively linear gradient are preferred to have an increased image quality. Based on the sparse Halbach cylinder proposed in$\,$\cite{cooley2015}, further optimization for a monotonic SEM was reported in$\,$\cite{cooley2018} using a genetic algorithm (GA) and finite-element simulations, to favor the nonlinear image reconstruction. 

There are other designs of magnets that are reported to generate linear SEMs in the literature. It is commonly seen in single-sided or unilateral NMR/MRI systems for one-dimensional profiling near the surface of a magnet.  
Examples in this category are a horseshoe magnet$\,$\cite{eidmann1996horseshoe_magnet,blumich2000horseshoe_magnet}, a magnet array consisting of many small magnet blocks with the optimized arrangement$\,$\cite{prado2003single,perlo2004single}, a single-bar magnet$\,$\cite{blumich2002bar_magnet}, a single magnet topped with a shaped iron pole cap$\,$\cite{marble2006constant}, and a Halbach cylinder when the stray field outside the cylinder is used$\,$\cite{chang2006single}. For the magnets reported for single-sided or unilateral NMR/MRI systems, although the SEMs are linear, the FoV is usually limited to the region close to the surface of the magnets, which is not large enough to accommodate and image a large sample (e.g., human head). A Halbach array allows a relatively large FoV when the inner field is used. Besides the optimization for a monotonic field reported in$\,$\cite{cooley2018}, in 2016, P. Bl\"{u}mler proposed to concentrically nested dipolar and quadrupolar Halbach cylinders, obtaining constant gradients for MRI imaging\,\cite{blumler2016proposal}. In\,\cite{blumler2016proposal}, the proposed nested Halbach cylinders theoretically shows an average field strength of 0.45\,T in a circular region with a diameter of 20$\,$mm, and the gradients in FoV can be varied from 44.5 to 53.0\,mT/m.

For a Halbach array, the magnetic field it supplies is in the transverse plane of the array cylinder, where the designs of traditional radiofrequency (RF) coils cannot be applied directly. Recently, a ring-pair magnet array was proposed based on an Aubert ring pair$\,$\cite{aubert1991,Ren2018}, showing a longitudinal field with a relatively high field strength and homogeneity in a cylindrical FoV (a diameter of 20 cm and a length of 5 cm)\,\cite{Ren2018}. The field pattern is concentric which intrinsically spatially encodes the NMR/MRI signal in the radial direction. For this design, the encoding along the $\theta-$direction is missing. To achieve a 2D spatial encoding, as suggested in\,\cite{Ren2018}, one method
is to apply coil sensitivity encoding provided by rotating receiving coils to obtain additional information in the $\theta-$direction\,\cite{pruessmann1999sense,Trakic2009Rotating_coil}. Alternatively, a rotating encoding magnet block (or blocks) can be introduced to break the axial symmetry of the magnetic field to bring additional encoding information in the $\theta-$direction for 2D imaging. 

In this paper, the Aubert ring-pair magnet array in$\,$\cite{Ren2018} is further optimized to obtain a 1D monotonic field pattern to favor image reconstruction, on top of having a relatively high field strength and the acceptable field inhomogeneity along the longitudinal direction. To achieve this goal, the ring pair is discretized into ring segment pairs for optimization, resulting in irregular-shaped rings. The resultant proposed design is named irregular-shaped ring-pair magnet array. A GA was applied and a current model for a ring segment pair (a fan-shaped pair) was derived and used for a fast forward calculation of the magnetic field in the optimization.   
The design and optimization are detailed in Section$\,$\ref{sec:methods}. The optimization results are presented in Section$\,$\ref{sec:results}, and the proposed magnet array is compared to a Halbach array when their fields are used for encoding for MR imaging numerically, to show that the proposed design favors image reconstruction. Discussions on the design and optimization are presented in Section$\,$\ref{sec:discussion}, and a conclusion is made in Section$\,$\ref{sec:conclusion}.

\section{Methods}\label{sec:methods}
\begin{figure}[t]
    \centering
      \subfloat[]{
        \includegraphics[height=0.22\linewidth]{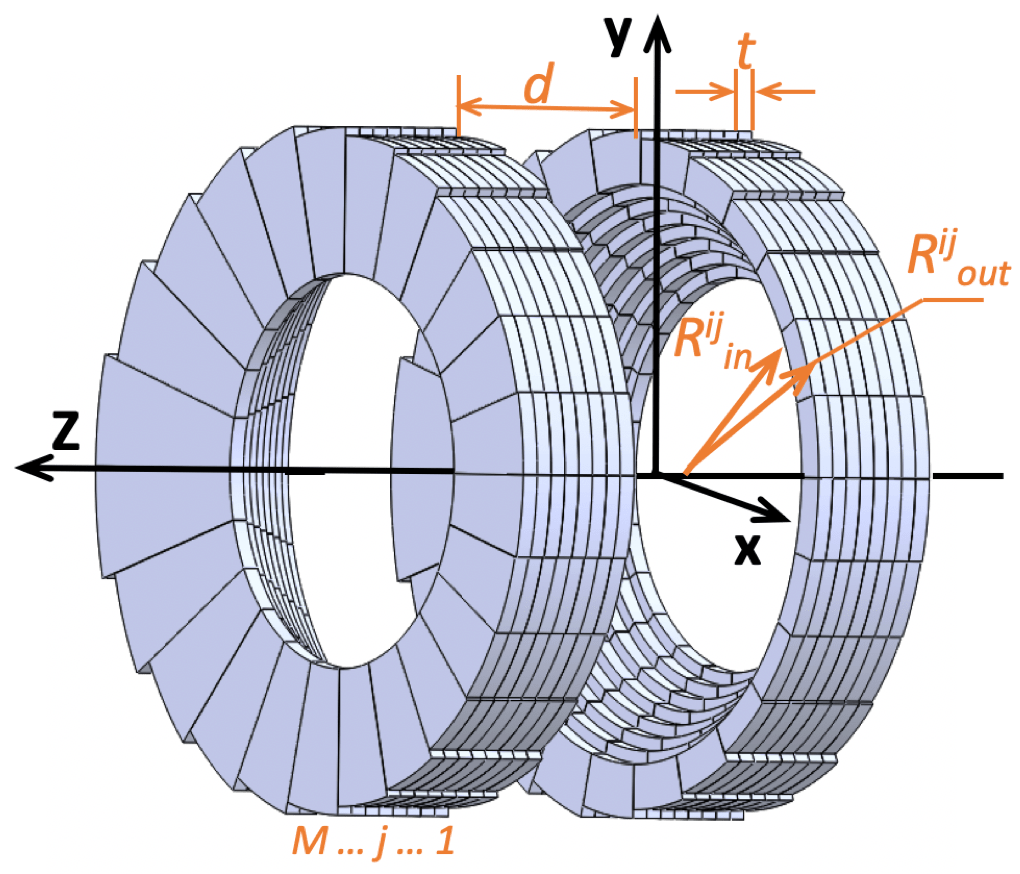}}
        \hspace{0.02\linewidth}
  \subfloat[]{
        \includegraphics[height=0.22\linewidth]{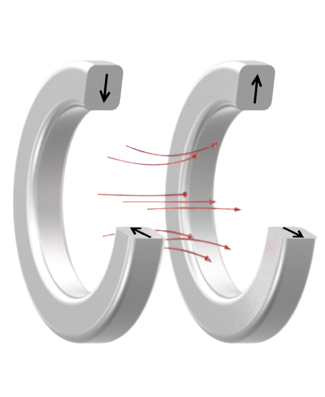}}
        \hspace{0.02\linewidth}
  \subfloat[]{
       \includegraphics[height=0.22\linewidth]{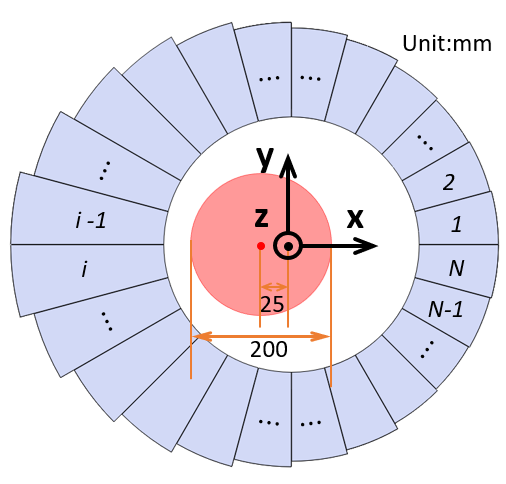}}
       \\
  \subfloat[]{
       \includegraphics[height=0.22\linewidth]{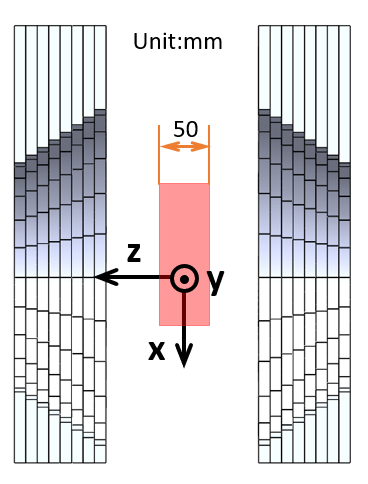}}
       \hspace{0.02\linewidth}
  \subfloat[]{
       \includegraphics[height=0.22\linewidth]{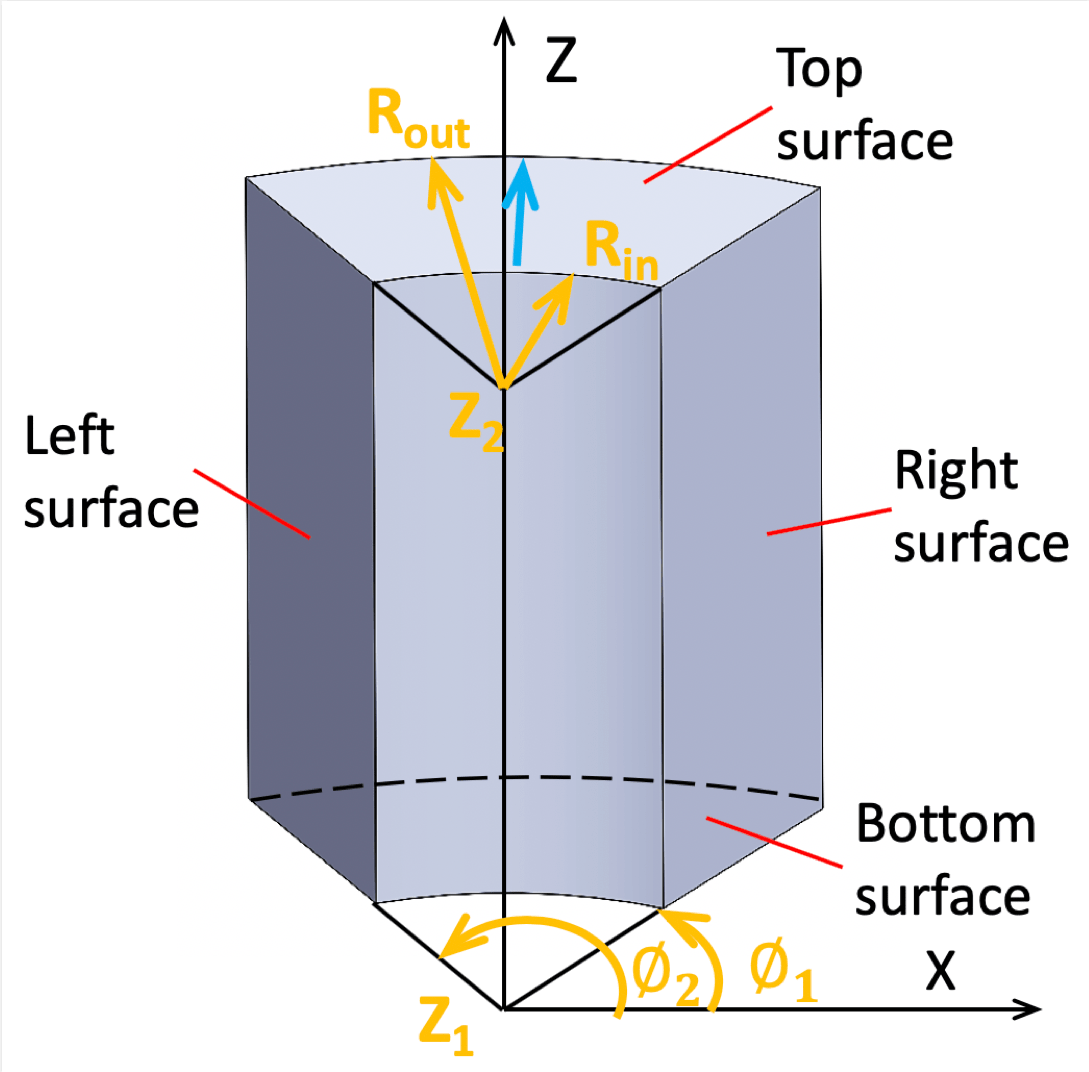}}
  \caption{(a) 3D view of the proposed irregular-shaped ring-pair magnet array. (b) The sectional view of an Aubert ring pair\,\cite{aubert1991}. (c) The front view of the proposed array, and the FoV is indicated in red. (d) The side cross sectional view of the proposed array. (e) One segmented fan-shaped magnet element with a radially outward polarization.}
\label{illustration of the proposed model}
\end{figure}

The proposed design is shown in Fig.$\,$\ref{illustration of the proposed model}$\,$(a) which consists of segmented Aubert ring pairs that have varying inner diameters from one pair to another, and different outer diameters from one segment to another. A basic Aubert ring pair is shown in Fig.$\,$\ref{illustration of the proposed model}$\,$(b)$\,$\cite{aubert1991}. 
As shown in Fig.$\,$\ref{illustration of the proposed model}$\,$(b), it consists of two annular magnets of the same dimension with the central axes aligned and located a distance apart, forming a cylindrical space. In the ring pair, one magnet ring has the magnetization radially pointing inward (the left one) and the other radially pointing outward (the right one). It supplies dipolar magnetic field along the axial direction of the cylinder (from left to right). 
In the proposed magnetic array as shown in Fig.$\,$\ref{illustration of the proposed model}$\,$(a), the proposed magnet array has $M$ magnet ring pairs and is symmetric with respect to the central plane ($xy-$plane, $z=0$). The rings in a pair are identical. All the rings have the same thickness $t$. Therefore, on each side, the aggregate consists of $M$ magnet rings. 
The distance between the two inner edges of the two inner most rings is $d$.
All the magnet rings on the left ($z>0$) are radially polarized outwards, and those on the right ($z<0$) are radially polarized inwards, thus a longitudinal magnetic field (along the $z-$axis) is generated. For each ring, they are segmented into $N$ fan-shaped segments, as shown in Fig.$\,$\ref{illustration of the proposed model}$\,$(a). Fig.$\,$\ref{illustration of the proposed model}$\,$(c) and (d) shows the front view and side sectional view of the proposed array. Fig.$\,$\ref{illustration of the proposed model}$\,$(e) shows a ring segment. For each segment, it is indexed using $j$ and $i$ where $j$ indexes the $j^{th}$ ring pair and $i$ indexes the $i^{th}$ ring segment. The inner and outer radius of each segment are denoted as $R_{in}^{ij}$ and $R_{out}^{ij}$, respectively.

For the optimization, the inner radii of the segments from the same ring pair, e.g. the $j^{th}$ ring pair, are set to be the same. Therefore, we let $R_{in}^{ij}=R_{in}^{j}$. On the other hand, the outer radii vary from one segment to another along a ring, while they are set to be the same across rings for the segments of the same index. For this reason, $R_{out}^{ij}$ is set to be $R_{out}^{i}$.   
In the optimization, $R_{in}^{j}$ and $R_{out}^{i}$ ($j =1,\dots, M$, and $i= 1,\dots, N$) are the geometrical parameters to be optimized to generate a targeted SEM with a desired field pattern, field strength, and homogeneity. The distance $d$ between the two inner edges of the two inner most rings is fixed to be 240\,mm. For head imaging in 2D, the FoV under optimization is set to be a cylindrical volume with a diameter of 200$\,$mm and a length of 50$\,$mm inside the magnet bore. It is 25$\,$mm off the origin along the $-x$-direction. It is illustrated in red in Fig.$\,$\ref{illustration of the proposed model}$\,$(c) and (d). 
For the optimization, GA was applied and a current model for the calculation of the magnetic field of  magnet ring segments was derived and used for a fast forward calculation. The details are presented next.

\subsection{Optimization using Genetic Algorithm}
GA$\,$\cite{GA1} was used for the optimization of the proposed magnet array. GA provides candidate solutions with a high diversity. Generally, it contains iterations with improvements where off-springs are produced, cross-overed, and mutated. The application of GA to the optimization of magnet system for MRI scanners has 
shown the effectiveness and efficiency of the tool$\,$\cite{shaw2002genetic, yao2005new, cooley2018,Ren2018}. The key of applying GA is the definition of an effective fitness function which rewards the good off-springs and penalizes the bad ones. In this study, there are three optimization objectives: a high average field strength ($>$ 100$\,$mT), the controlled field inhomogeneity ($<$10$\,$mT), and a monotonic field pattern that is as linear as possible. 
All three objectives are combined in a single fitness function to accelerate the optimization rather than using a multi-objective GA. This fitness function is shown as follows, 
\begin{equation}\label{fitness_function}
\begin{split}
 \min\,f
 &=\bigg\|\frac{\max_{1\leq k \leq N_1}\bm{B}_z(\bm{r}_k)-\min_{1\leq k \leq N_1}\bm{B}_0(\bm{r}_k)}{\mathrm{mean}\big(\bm{B}_z(\bm{r}_k)\big)}\bigg\|\times 10^6-\alpha\big(\sum_{k=1}^{N_1} \bm{B}_z(\bm{r}_k)/N_1-100\big)\\
 &+\beta\big(N_2-\sum_{h=1}^{N_2} \mathrm{issorted}([\bm{B_z({r}^1_{h})},\cdots,\bm{B_z({r}^\ell_h)},\cdots,\bm{B_z({r}^{N_3^h}_h)}])\big)
\end{split}
\end{equation}
where, $k$ is the index of the observation points $\bm{r}$ in the FoV and $N_1$ is the total number of the points, $N_2$ is the total number of the observation lines parallel with the $x-$axis in FoV with a spatial step of 10$\,$mm, $h$ is the index of the observation lines in FoV, $\bm{r}^\ell_h$ is the $\ell^{th}$ point along the $h^{th}$ observation line. The total number of points along the $h^{th}$ line is denoted using $N_3^h$.
The first term in (\ref{fitness_function}) is the field inhomogeneity, and the unit is part per million (ppm). The second term in (\ref{fitness_function}) rewards the off-springs with a field strength higher than 100\,mT, and penalizes those with a field strength lower than 100\,mT.  
The third term in (\ref{fitness_function}) takes care of the monotonicity of the field pattern in FoV by checking the field gradient along the $x-$direction. In our approach, function $\textit{issorted}$ was used for testing the monotonicity of the field along the observation line, and it takes less computation compared to calculating the number of voxels with undesired gradients$\,$\cite{cooley2018}.
To balance the three objectives, weighting coefficients $\alpha$ and $\beta$ were applied to the second and third terms in (\ref{fitness_function}), respectively. In this study, it was empirically set to be 5$\times10^{3}$ and 2.5$\times10^{4}$, respectively.  

\subsection{Forward Calculation: Current Model} 
As known in the literature, the current model is suitable to calculate the magnetic field of a yokeless magnet system without the segmentation of the magnets\,\cite{furlani2001current_model}. The calculation using a current model is more efficient compared to finite element method (FEM)\,\cite{chari1980FEM}, or Boundary Integral Method (BIM)\,\cite{elleaume1997BIM}. For the proposed magnet array, a current model for a fan-shaped ring segment (shown in Fig.\,\ref{illustration of the proposed model}\,(e)) was derived and implemented for a fast forward calculation for the optimization. The details of the derivation are shown next. In the derivation, a cylindrical coordinate system was used. 

\begin{figure}[t]
    \centering
      \subfloat[]{
        \includegraphics[height=0.22\linewidth]{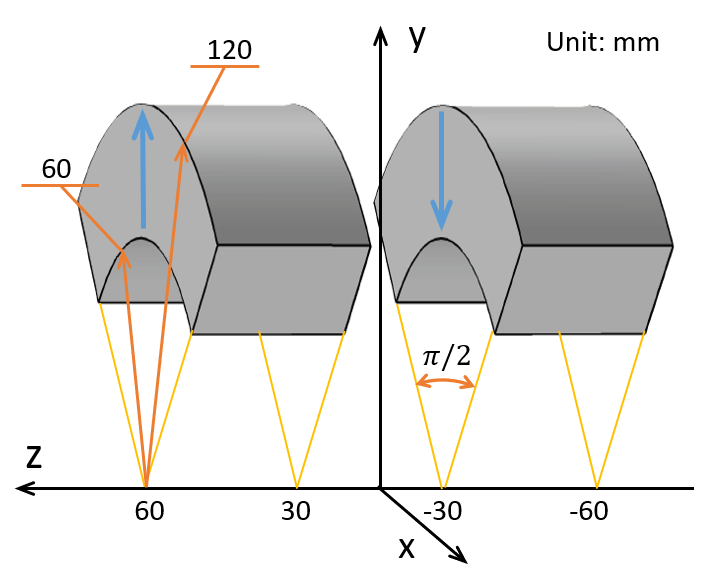}}
        \hspace{0.02\linewidth}
  \subfloat[]{
        \includegraphics[height=0.22\linewidth]{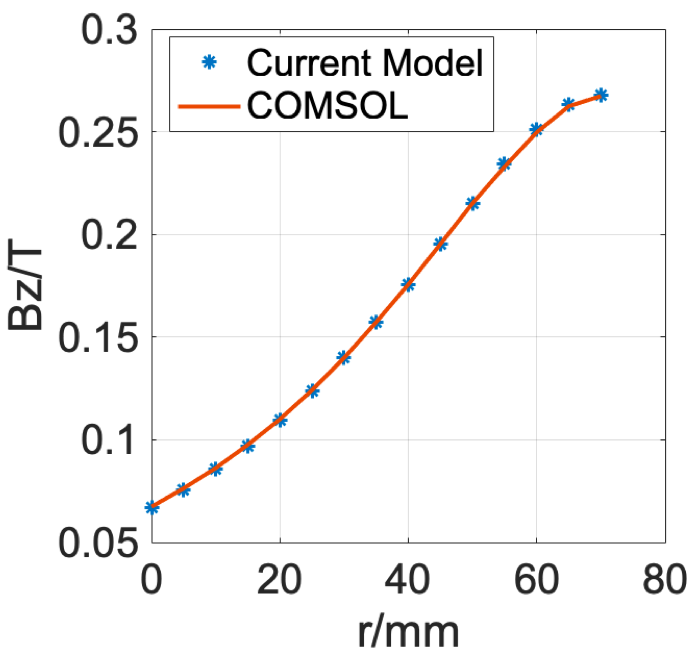}}
  \caption{(a) The 3D view of a fan-shaped magnet segment pair (the blue arrows indicate the polarization of the magnets), (b) The calculated $\bm{B_z}$ on the observation points along the y-axis (from 0 to 70$\,$mm with a step of 5$\,$mm) based on (\ref{eq3}) using MATLAB and those using COMSOL Multiphysics. }
  \label{fig_validation_of_CM}
\end{figure}

In a current model, a permanent magnet is modeled using equivalent current sources. The magnetic field generated by these equivalent current sources can be calculated by (\ref{eq_o}) below, 
\begin{align} \label{eq_o}
\begin{split}
\bm{B}(\bm{r})=\frac{\mu_0}{4\pi} \oint_{S} \bm{j_m}(\bm{r}^{\prime}) \times \frac{\bm{r}-\bm{r}^{\prime}}{|\bm{r}-\bm{r}^{\prime}|^3} d{s'}+\frac{\mu_0}{4\pi} \oint_{V} \bm{J_m}(\bm{r}^{\prime}) \times \frac{\bm{r}-\bm{r}^{\prime}}{|\bm{r}-\bm{r}^{\prime}|^3} d{v'}
\end{split}
\end{align}
where, $\mu_0$ is the permeability of free space, $\bm{r}=\left<r,\phi,z\right>$ is the observation point, $\bm{r}^{\prime}=\left<r^\prime,\phi^\prime,z^\prime\right>$ is current source point, $\bm{j_m}$ is the equivalent surface current source, $\bm{J_m}$ is the equivalent volume current source, and $S$ and $V$ are the surface and the volume of the magnet, respectively. For a fan-shaped magnet shown in Fig.$\,$\ref{illustration of the proposed model}$\,$(e), the equivalent volume current density is determined by
\begin{equation}\label{volume_current_density}
    \bm{J_m} = \nabla\times\bm{M_0} =0
\end{equation}
where, $\bm{M_0}$ is the magnitude of the remanent magnetization of a permanent magnet. As the remanent magnetization is a constant along different polarization, its curl is zero. Thus, based on (\ref{volume_current_density}), the equivalent volume current is zero. Therefore, in (\ref{eq_o}), there are only equivalent surface currents remained. (\ref{eq_o}) is rewritten as,
\begin{align} \label{eq_o_short}
\begin{split}
\bm{B}(\bm{r})=&\frac{\mu_0}{4\pi} \oint_{S} \bm{j_m}(\bm{r}^{\prime}) \times \frac{\bm{r}-\bm{r}^{\prime}}{|\bm{r}-\bm{r}^{\prime}|^3} d{s'}
\end{split}
\end{align}
For the equivalent surface currents, they are expressed as,
\begin{equation}\label{current_source}
\bm{j_m}(\bm{r}) = \bm{M_0}\times\hat{n}=
\begin{cases}
-M_0 \hat{{\phi}^{\prime}} & \mathrm{top\; surface} \\
M_0 \hat{{\phi}^{\prime}} & \mathrm{bottom\; surface}\\
-M_0\hat{z^{\prime}} & \mathrm{left\; surface}\\
M_0\hat{z^{\prime}} & \mathrm{right\; surface}
\end{cases}
\end{equation}
where, $\hat{n}$ denotes the unit normal vector of the magnet surface. Substituting (\ref{current_source}) into (\ref{eq_o_short}) gets an expression for $\bm{B}(\bm{r})$. The $z-$component of $\bm{B}(\bm{r})$ dominates the magnetic field generated by the proposed magnet array, and $\bm{B}_{z}(\bm{r})$ generated by the fan-shaped magnet ring segment is expressed as,
\begin{align}\label{eq3}
\begin{split}
 \bm{B}_{z}(\bm{r})  = \frac{-\mu_0 M_0}{4\pi} \oint_{S_{top}}\frac{C_1}{C_2}ds^{\prime} +\frac{\mu_0 M_0}{4\pi} \oint_{S_{bottom}}\frac{C_1}{C_3}ds^{\prime}
\end{split}
\end{align}
where
\begin{align}\label{eq4}
\begin{split}
C_1 & = -r cos(\phi - \phi^\prime) + \phi sin(\phi - \phi^\prime) + r^\prime\\
C_2 & = \big(r^2+{r^{\prime}}^2-2rr^{\prime}cos(\phi-\phi^{\prime})+(z_1-z')^2)\big)^{(3/2)}\\
C_3 & = \big(r^2+{r^{\prime}}^2-2rr^{\prime}cos(\phi-\phi^{\prime})+(z_2-z')^2)\big)^{(3/2)}
\end{split}
\end{align}
The superposition principle holds in a yokeless magnet system, so the total resultant magnetic field $\bm{B_{total}}(\bm{r})$ generated by the whole proposed magnet array in Fig.\,\ref{illustration of the proposed model}\,(a) can be calculated by
\begin{align}\label{eq_BTotal}
\begin{split}
\bm{B_{total}}(\bm{r})= \sum_{j=1}^{M}\sum_{i=1}^{N} \big[ \bm{B}(\bm{r},R_{in}^j,R_{out}^i,z_{1}^{j},z_{2}^{j},\phi_{1}^{i},\phi_{2}^{i})-\bm{B}(\bm{r},R_{in}^j,R_{out}^i,-z_{1}^{j},-z_{2}^{j},\phi_{1}^{i},\phi_{2}^{i})\big]
\end{split}
\end{align}
To validate the derivation, a fan-shaped segment pair shown in Fig.$\,$\ref{fig_validation_of_CM}$\,$(a) was calculated using both (\ref{eq_BTotal}) and COMSOL Multiphysics. Fig.$\,$\ref{fig_validation_of_CM}$\,$(b) shows both results along the radial direction. As shown in Fig.$\,$\ref{fig_validation_of_CM}$\,$(b), the result using the derived current model and  that using COMSOL Multiphysics (FEM-based) show a good agreement with each other. 

In the GA optimization, the number of ring pairs, $M$, was set to be 9, the thickness of each ring, $t$, was set to be 12\,mm, the distance of the two innermost rings, $d$, is set to be 240\,mm. The remanence of all the magnets were set to be 1.4$\,T$. 
Based on the report on an optimization of an Aubert ring pair aggregate in\,\cite{Ren2018}, a relatively high field strength with the low inhomogeneity can be obtained when the inner radii of the ring pairs are tapered outside in. To accelerate the optimization, in the current optimization, the inner radii of the ring pairs, $R^j_{in}$ ($j=1,2,\ldots,9$), were tapered outside in in the initial candidate solutions. Moreover, each magnet ring was segmented into 24 fan-shaped segments, resulting in 24 $R_{out}^{i}$ ($i=1,2,\ldots,24$) in one individual for optimization. To further accelerate the optimization efficiency, the symmetry with respect to the central $r\theta-$plane ($xy-$plane) and to the $x-$axis was set, and this reduce the parameters to be optimized to be 12, which are $R_{out}^{i}$ ($i=1,2,\ldots,12$) in one individual. 
 
To have a monotonic field along the x-direction in the FoV, tapered outer radii along a semi-ring are helpful. Therefore, $R_{out}^{i}$ ($i=1,2,\ldots,12$) were set to be tapered and the number of optimization parameters was further reduced. The governing equation as follows are imposed to $R_{out}^{i}$'s,
\begin{equation}\label{eq:tapper_func}
R_{out}^{i} = R_{out}^{max}-\rho(i-1)^\sigma \quad i=1,2,\dots,12
\end{equation}
where, $\rho$ is the step reduction of the outer radius as the index of $i$ increases, $\sigma$ is the order of the tapering function, and $R_{out}^{max}$ denotes the maximum radius among $R_{out}^{i}$ ($i=1,2,\ldots,12$) under optimization. With the tapering function in  (\ref{eq:tapper_func}), the number of parameters in one individual is reduced from 12 to only 3. In the GA optimization, $\rho \in [0.1\;4]$, $\sigma \in [0.1\;5]$, $R_{out}^{max}\in[290\;320]$, and the population size was set to be 50 which can provide enough diversity for the candidate solutions. With the current model (the calculation time is 1/10 of that using FEM)  and the reduced number of optimization parameters, the forward calculation for one iteration was greatly accelerated.
\begin{figure}[ht]
    \centering
        \subfloat[]{
        \includegraphics[height=0.22\linewidth]{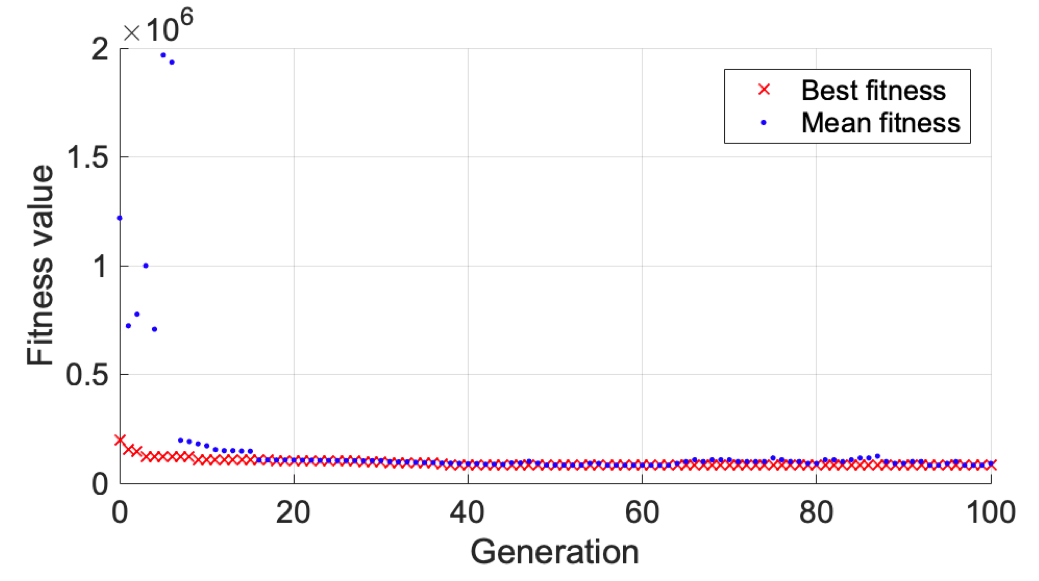}}
        \hspace{0.02\linewidth}
        \subfloat[]{
        \includegraphics[height=0.22\linewidth]{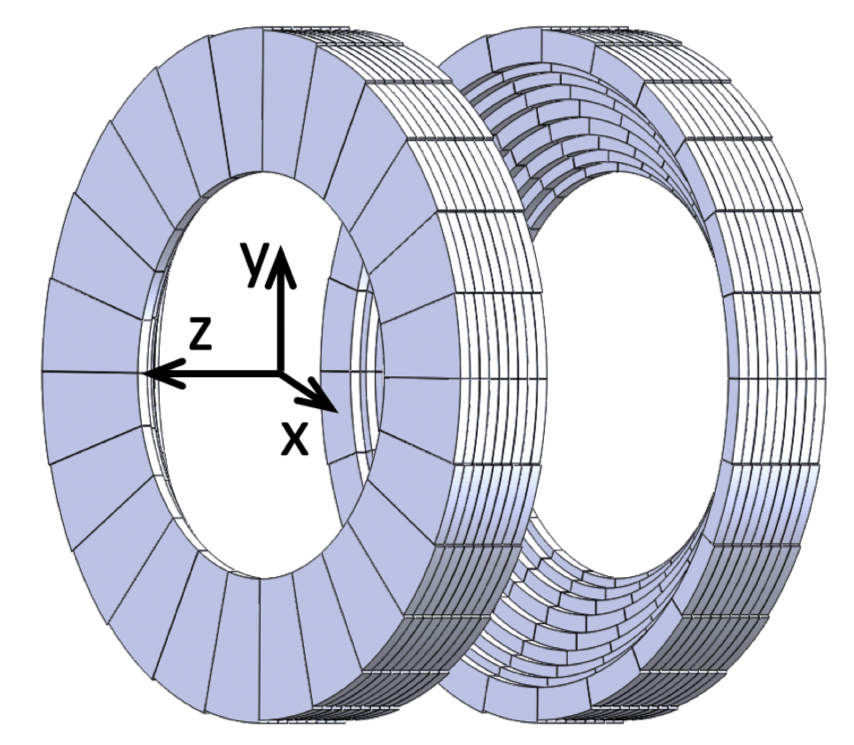}}
        \hspace{0.02\linewidth}
        \subfloat[]{
        \includegraphics[height=0.22\linewidth]{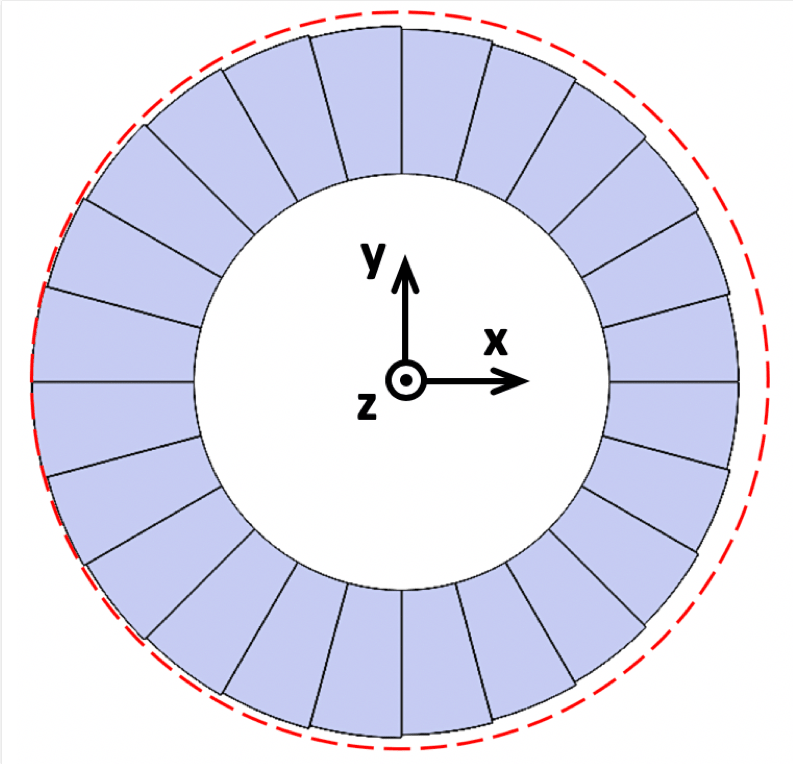}}
  \caption{(a) The change of fitness value versus iteration steps. (b) 3D model of the optimal design. (c) Front view  of the optimal magnet array is shown. The red dashed circle is a reference circle with a radius of 320\,mm.}
\label{optimization_result}
\end{figure}
\begin{figure}[t]
    \centering
        \subfloat[]{
        \includegraphics[height=0.2\linewidth]{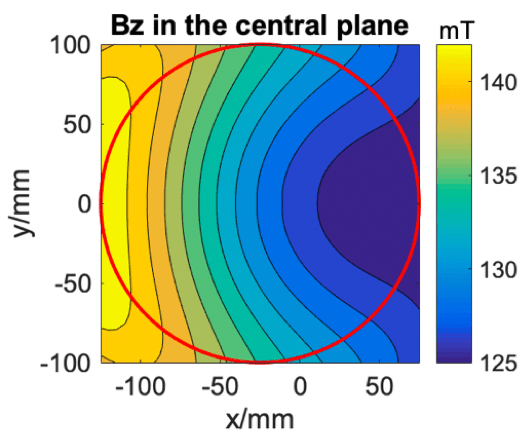}}
        \hspace{0.02\linewidth}
        \subfloat[]{
        \includegraphics[height=0.2\linewidth]{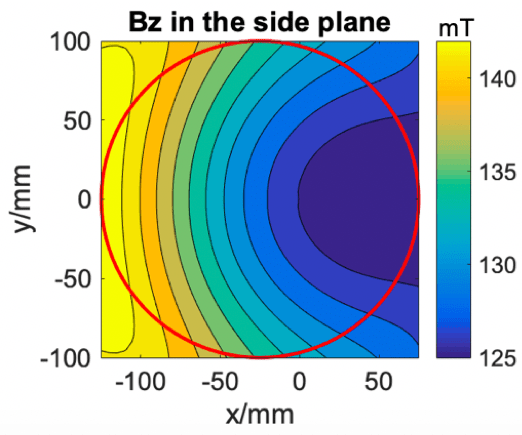}}
        \hspace{0.02\linewidth}
        \subfloat[]{
        \includegraphics[height=0.2\linewidth]{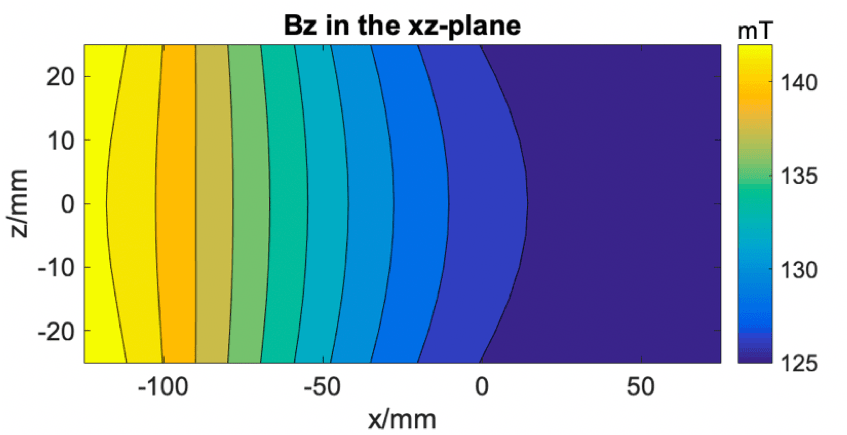}}
        \\
        \subfloat[]{
        \includegraphics[height=0.2\linewidth]{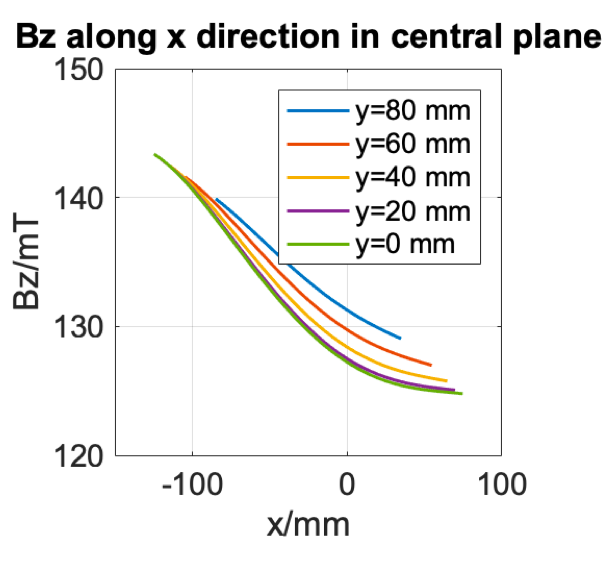}}
        \hspace{0.02\linewidth}
        \subfloat[]{
        \includegraphics[height=0.2\linewidth]{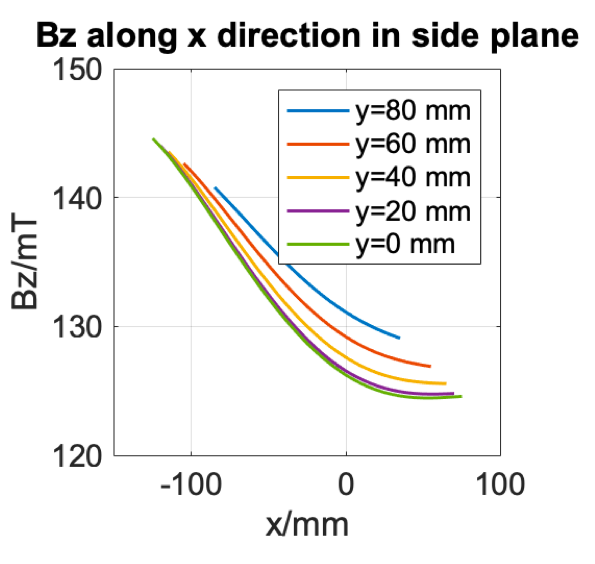}}
  \caption{The $\bm{B_z}$ generated by the optimized magnet array (a) in the central $xy-$plane (z\,=\,0\,mm), (b) in the side $xy-$plane (z\,=\,25\,mm), and (c) in the $xz-$plane (y\,=\,0\,mm) of the FoV calculated in Matlab; the $\bm{B_z}$ along the $x-$direction at y = 0, 20, 40, 60, 80\,mm (d) in the central $xy-$plane (z\,=\,0\,mm), and (e) in the side $xy-$plane (z\,=\,25\,mm) in the FoV.}
\label{field_plot_of_optimization_result}
\end{figure}

\section{Results}\label{sec:results}
During the optimization process, the optimization was repeated multiple times since GA did not always converge to the same result. Fig.\,\ref{optimization_result}\,(a) shows that both the mean and the best fitness value decreased as the number of iterations increased in one optimization loop. A good result with a fitness value of 61584 is presented here with a trade-off among the field strength, field inhomogeneity, and monotonicity of field pattern along the x-direction. The preset $R^j_{in}$ ($j=1,2,\ldots,9$) were [276.0 267.6 256.8 243.6 228.0 212.4 199.2 188.4 180.0] (unit:\,mm), and the optimized values of $\rho$, $\sigma$, and $R_{out}^{max}$ were 2.3, 1.05 and 320\,mm, respectively. Based on (\ref{eq:tapper_func}), the optimized $R^i_{out}$ ($i=1,2,\ldots,12$) were [320.0 317.7 315.2 312.7 310.1 307.5 304.9 302.3 299.6 296.9 294.2 291.5] (unit:\,mm). The optimized array was modelled in SolidWorks according to the optimized parameters, and it is shown in Fig.\,\ref{optimization_result}\,(b) and (c). 

The field patterns of the optimal structures were calculated using (\ref{eq3}) for an evaluation. The optimized magnet array provides a $\bm{B_0}$ field with an average field strength of 132.98$\,$mT and the field homogeneity of 151840$\,$ppm in the FoV. Fig.\,\ref{field_plot_of_optimization_result}\,(a-c) show the $B_z$ distribution in the $xy-$plane at z = 0 and 25\,mm, and $B_z$ distribution in the $xz-$plane at y\,=\,0\,mm, respectively. Fig.\,\ref{field_plot_of_optimization_result}\,(d) and (e) show the $B_z$ along the $x-$direction at y = 0, 20, 40, 60 and 80\,mm within the planes shown in Fig.\,\ref{field_plot_of_optimization_result}\,(a) and (b) in the FoV, respectively. 
As shown in Fig.\,\ref{field_plot_of_optimization_result}\,(a), (b) and (c), $B_z$ decreases monotonically from left to right, especially when -125\,mm\,$<x<$\,25\,mm. The monotonicity of the field can clearly be seen from the 1D plots along different lines in Fig.\,\ref{field_plot_of_optimization_result}\,(d) and (e). 
When -125\,mm\,$<x<$\,50\,mm and -20\,mm\,$<y<$\,20\,mm, the gradient is about 122.5\,mT/m and 140\,mT/m in the $xy-$plane at z\,=\,0 and 25\,mm, respectively. Outside the region of -20\,mm\,$<y<$\,20\,mm, the region that shows linear fields is smaller and the gradient is smaller as well. For example, when y = 60\,mm, the field shows a linearity from -105\,mm to 0\,mm, with a gradient of about 105.7\,mT/m and 121\,mT/m in the $xy-$plane at z\,=\,0 and 25\,mm, respectively. As can be seen from Fig.\,\ref{field_plot_of_optimization_result}\,(a)-(e), through the optimization when the inner radii of the rings were tapered outside in and the outer radii of each ring segment were tapered along the x-direction and optimized, a field monotonicity along the x-axis was successfully obtained for imaging.

\begin{figure}[ht]
    \centering
    \subfloat[]{
        \includegraphics[height=0.22\linewidth]{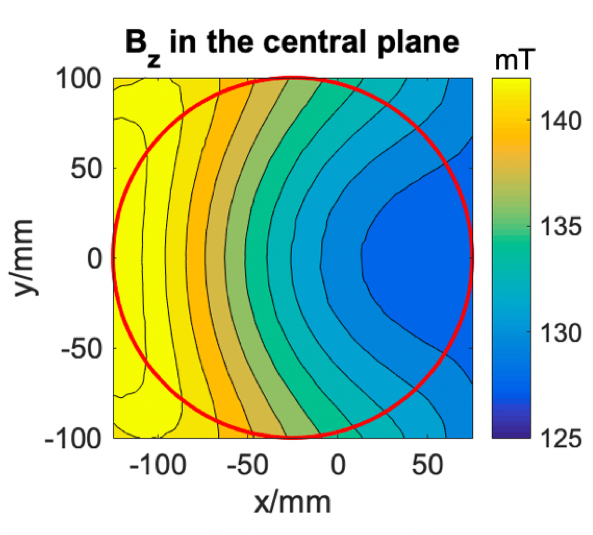}}
        \hspace{0.02\linewidth}
    \subfloat[]{
        \includegraphics[height=0.22\linewidth]{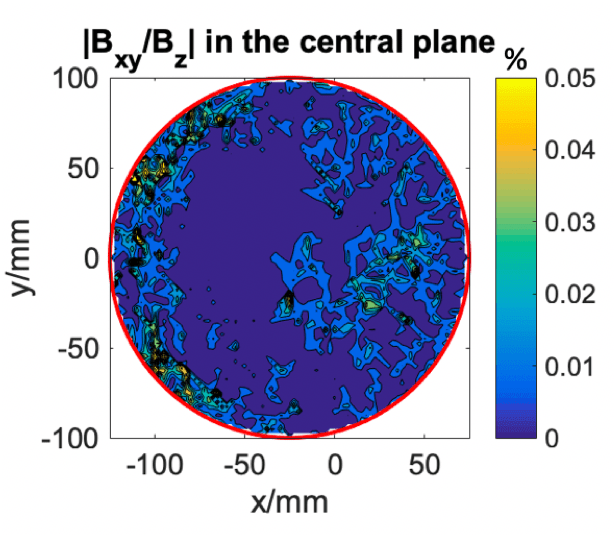}}
        \hspace{0.02\linewidth}
    \subfloat[]{
        \includegraphics[height=0.22\linewidth]{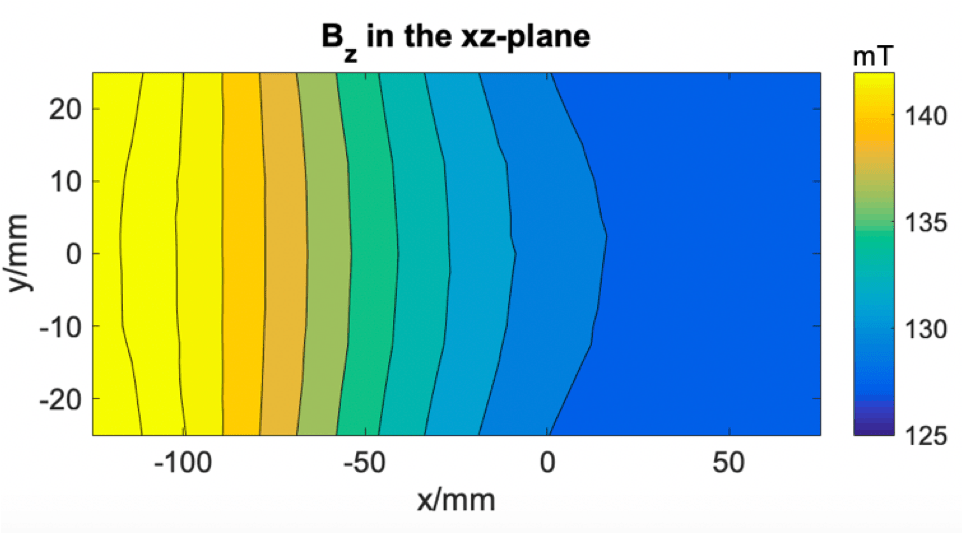}}
  \caption{(a) The simulated $\bm{B_0}$ field in the central $xy-$plane (z\,=\,0\,mm). (b) $|\bm{B_{xy}}/\bm{B_0}|$ (unit:$\,\%$) in central plane of the FoV. (c) The simulated $\bm{B_0}$ field in the $xz-$plane (y\,=\,0\,mm).}
\label{fig:comsol_validation}
\end{figure}

A realistic simulation was done in COMSOL Multiphysics to validate the optimal design. The simulated $\bm{B_z}$ field in the central $xy-$plane (z\,=\,0\,mm) and in the $xz-$plane (y\,=\,0\,mm) were shown in Fig.\,\ref{fig:comsol_validation}\,(a) and (c), respectively. Compared to the $\bm{B_0}$ field shown in Fig.\,\ref{field_plot_of_optimization_result}\,(a) and (c), the current model and COMSOL Multiphysics showed the almost same field pattern in the FoV for the optimal design. The average field strength and field inhomogeneity from the COMSOL Multiphysics were 134.69\,mT and 145680\,ppm, and the differences compared to the current model are within 1.3\,\% and 4.1\,\%, respectively. They showed good agreement with each other, and the effectiveness of the optimization was validated. The $x-$ and $y-$components of the magnetic field in FoV were also studied here, and the $|\bm{B_{xy}}/\bm{B_0}|$ (unit:$\,\%$) in the FoV was calculated and shown in Fig.\,\ref{fig:comsol_validation}\,(b). As can be seen, $|\bm{B_{xy}}/\bm{B_0}|$ is below 0.05\,$\%$ in the FoV, and the $z-$components are dominant in the $\bm{B_0}$ field generated. 

\begin{figure}[t]
    \centering
      \subfloat[]{
        \includegraphics[height=0.22\linewidth]{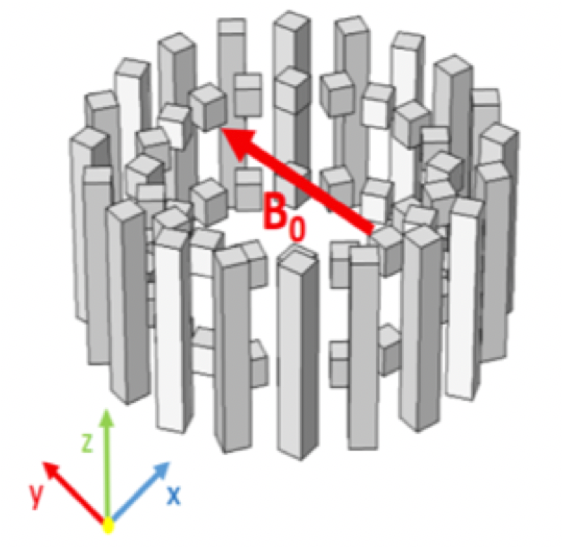}}
        \hspace{0.02\linewidth}
  \subfloat[]{
        \includegraphics[height=0.22\linewidth]{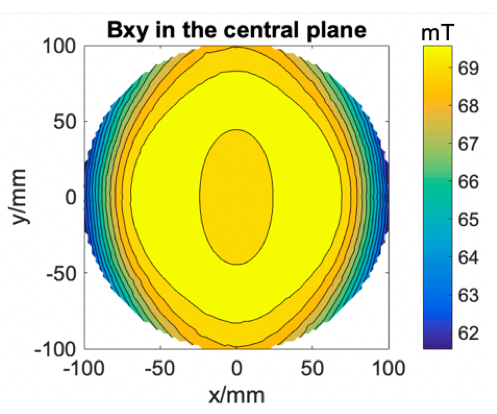}}
  \caption{(a) 3D model of a reference short sparse Halbach array\,\cite{ren2015}, the outer diameter of the Halbach cylinder is 380\,mm, and it consists of 20 1'$\times$1'$\times$6.5' N52 NdFeB magnet bars and 40 1'$\times$1'$\times$1' N52 NdFeB magnet cubes. (b) The simulated field distribution in the central $xy-$plane (z\,=\,0\,mm) of the Halbach array using COMSOL Multiphysics.}
\label{halbach_array}
\end{figure}

The optimized magnet array is compared to a sparse Halbach array in terms of the fields they generate, and when the fields they generate are used as SEM's for imaging.
Fig.\,\ref{halbach_array}\,(a) and (b) show a 3D model of a short sparse Halbach array reported in\,\cite{ren2015}, and the simulated field distribution in the central $xy-$plane (with a diameter of 200\,mm), respectively. As can be seen in Fig.\,\ref{halbach_array}\,(b), there are considerable regions where gradients are low or zero, especially in the central region.
Comparing the field pattern of the proposed magnet array in Fig.\,\ref{field_plot_of_optimization_result}\,(a-c) and that of the Halbach array in Fig.\,\ref{halbach_array}\,(b), it can be seen that most of the regions with low or zero gradients were successfully eliminated. 

\begin{figure}[ht]
    \centering
    \subfloat[]{
        \includegraphics[height=0.19\linewidth]{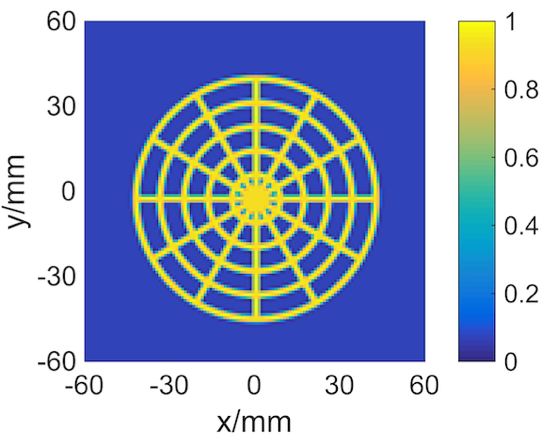}}
        \hspace{0.001\linewidth}
         \subfloat[]{
        \includegraphics[height=0.19\linewidth]{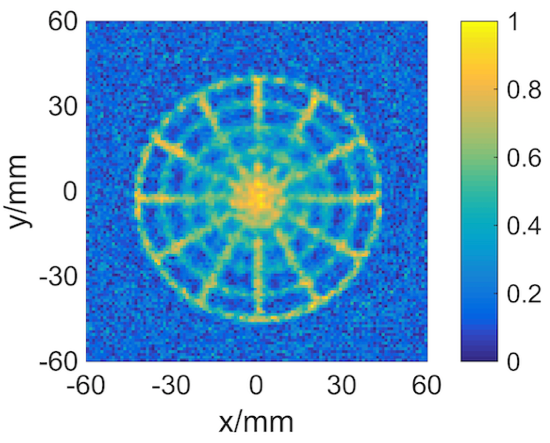}}
        \hspace{0.001\linewidth}
    \subfloat[]{
        \includegraphics[height=0.19\linewidth]{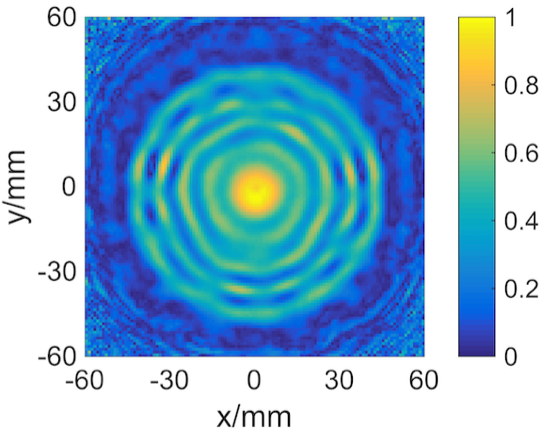}}
        \hspace{0.001\linewidth}
    \subfloat[]{
        \includegraphics[height=0.19\linewidth]{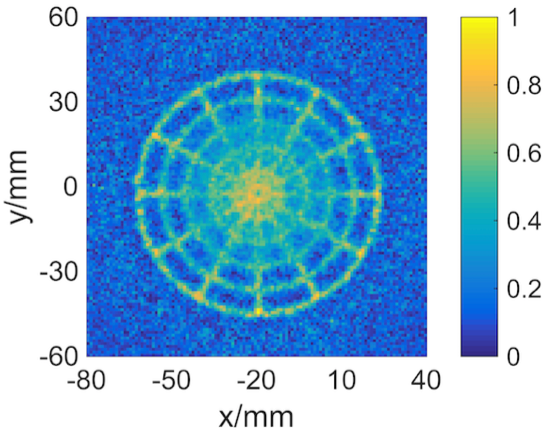}}
  \caption{(a) A web-pattern 2D phantom. (b) Reconstructed image using the $\bm{B_0}$ field generated by the optimized magnet array. (c) Reconstructed image using the reference magnetic field generated by the Halbach array in Fig.\,\ref{halbach_array}. (d) Reconstructed image using the reference magnetic field generated by the further optimized array with improved monotonicity in Fig.\,\ref{fig:linear_one}.}
\label{reconstructed_images}
\end{figure}

The fields from both the proposed array and the Halbach array (shown in Fig.\,\ref{field_plot_of_optimization_result}\,(a) and Fig.\,\ref{halbach_array}\,(b) were evaluated by examining the quality of reconstructed images when they are applied as SEM to encode signals of a web-pattern phantom shown in Fig.\,\ref{reconstructed_images}\,(a) numerically. The web pattern was used here to clearly show the resolution of the reconstructed image. 
For encoding, both SEMs were rotated $360^{\circ}$ at a step of $10^{\circ}$. Eight surface coils with a diameter of 3\,cm were located around the side wall of the cylindrical VoF with a diameter of 20\,cm, and used for signal reception. At each angle, 512 readouts were collected for image reconstruction and the SNR of signal was set to be 70\,dB.
Fig$\,$\ref{reconstructed_images}\,(b), and (c) shows the numerically reconstructed images using the central part ($120\times120\,mm^2$) of of the magnetic field from the proposed array (Fig.\,\ref{field_plot_of_optimization_result}\,(a)), and that using the field of the Halbach array (Fig.\,\ref{halbach_array}\,(b)), respectively. Kaczmarz iteration method was used for the reconstruction. Comparing the images in Fig$\,$\ref{reconstructed_images}\,(b) and (c), clearer features can be seen in the image by using the fields from the proposed array which have an improved monotonicity. The normalized root mean square error (NRMSE) of the image using the proposed array was reduced by about $20\,\%$ from $30.07\,\%$ to $24.21\,\%$ compared to that using the reference Halbach array. 
As can be seen in Fig$\,$\ref{reconstructed_images}\,(b), the blurring, especially at the center of the image is greatly reduced with the proposed array compared to the reference Halbach array. The structural similarity index (SSIM) of the image using the proposed array is 0.446, which is an improvement of about 40\% from that using the Halbach array (the SSIM is 0.321). Clearer images can be reconstructed using the magnetic field of the proposed array, which is attributed to the removal of the central zero-gradient region, and that of the regions with low gradients in the field pattern of the proposed array from that of a Halbach array.

\section{Discussion}\label{sec:discussion}
It has been shown that the proposed irregular-shaped ring-pair magnet array provides a monotonic field along a single direction in the central plane of a cylindrical FoV that works for head imaging. Through optimization using GA, the field strength is maintained at above 130\,mT with the inhomogeneity controlled at 151840\,ppm within the FoV. Due to the linearity in gradient, it serves better as a SEM leading to a better image quality in an MRI system, comparing to a Halbach array\,\cite{cooley2015,ren2017} and an Aubert-ring-pair aggregate\,\cite{Ren2018}. For a Halbach array with a quadrupolar $\bm{B_0}$ field pattern, there are regions where the gradients are low or zero, whereas for an Aubert-ring-pair aggregate\,\cite{Ren2018}, a gradient in the $\theta-$direction is missing although they can be compensated by adding additional structures or additional mechanical movements of the magnet when conducting encoding.
Comparing to an optimized sparse Halbach array\,\cite{cooley2018}, the proposed magnet array has a similar length (45.6\,cm vs. 45.7\,cm in length), but a larger diameter by 21.9$\%$ (64\,cm vs. 52.5\,cm in diameter), and both were designed and optimized to obtain a monotonic SEM for imaging in a FoV with a diameter of 20\,cm. The proposed array generates a $\bm{B_0}$ field with an average field strength of 132.98\,mT, which is 64\,\% higher compared to 81.1\,mT of the sparse array. However, the field inhomogeneity of the sparse Halbach array is only about 10$\,\%$ of the proposed array. A large inhomogeneity requires a wider bandwidth of the RF system, which will be discussed in the next paragraph. Overall, the proposed irregular-shaped ring-pair magnet array can be a good candidate to supply the SEM in a low-field portable MRI system, besides a sparse Halbach array\,\cite{cooley2018}.

The field homogeneity from the optimized array is decreased from 24786\,ppm to 151840\,ppm compared to the Aubert-ring-pair aggregate in$\,$\cite{Ren2018}, which can be considered as the price paid for the field monotonicity to favor spatial encoding. The low homogeneity results in the fact that a wide bandwidth are needed for radio-frequency (RF) excitation and signal reception, which may require RF coils and the CONSOLE to have wide working frequency bands. Therefore, as a result, the RF hardware has to be re-engineered to obtain a corresponding broad working bandwidth. Ultra broadband techniques for RF system has been proposed for NMR/MRI systems\,\cite{song2007,song2014,prado2003single}. A broad-band RF system can by achieved by using, for example, a transformer-based non-resonant transmitter and receiver which covers a broad frequency range$\,$\cite{song2007,song2014}, or using a receiver with a binary switch array to switch the resonant frequencies spanning in a significantly wide bandwidth\,\cite{prado2003single,song2007,song2011binary_switch}.

\begin{figure}[t]
    \centering
    \subfloat[]{
        \includegraphics[height=0.2\linewidth]{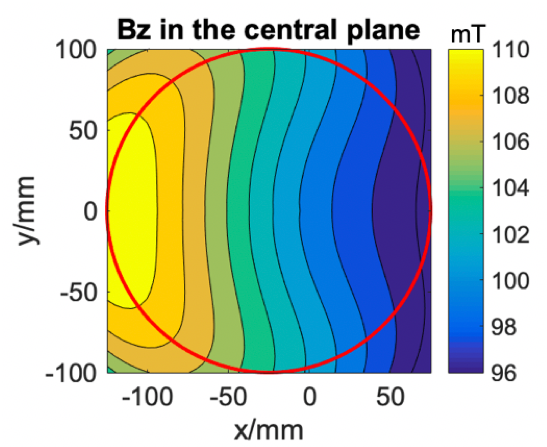}}
        \hspace{0.02\linewidth}
    \subfloat[]{
        \includegraphics[height=0.2\linewidth]{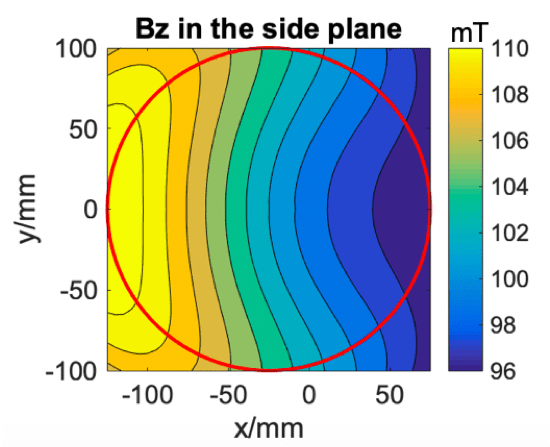}}
        \hspace{0.02\linewidth}
          \subfloat[]{
        \includegraphics[height=0.2\linewidth]{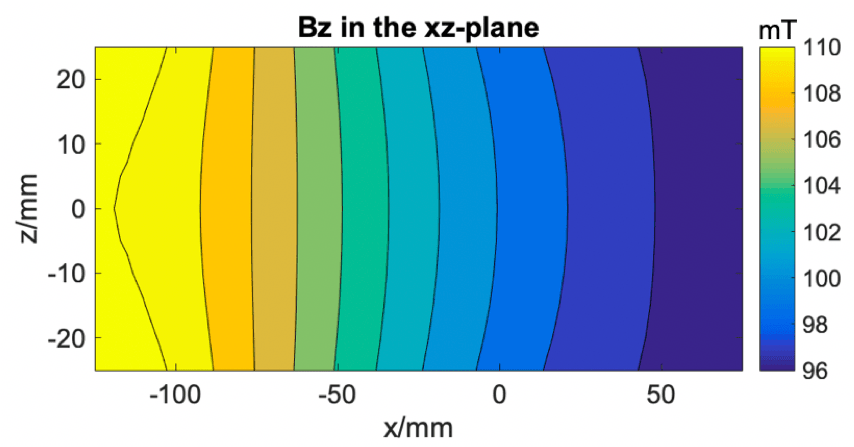}}
        \\
    \subfloat[]{
        \includegraphics[height=0.2\linewidth]{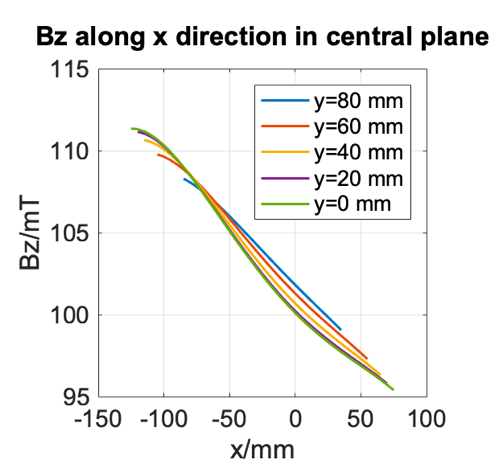}}
        \hspace{0.02\linewidth}
    \subfloat[]{
        \includegraphics[height=0.2\linewidth]{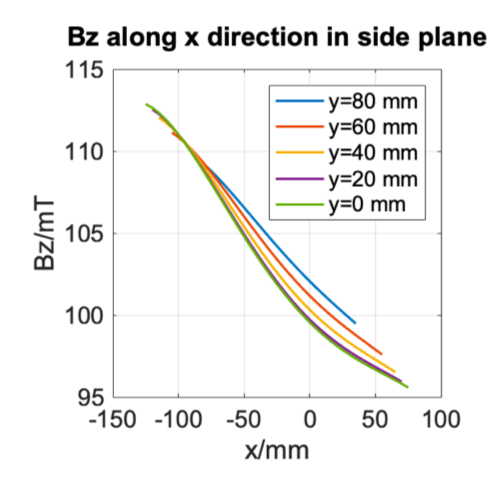}}
  \caption{ The $\bm{B_z}$ generated by a further optimized design with more field monotonicity (a) in the central $xy-$plane (z\,=\,0\,mm), (b) in the side $xy-$plane (z\,=\,25\,mm), and (c) in the $xz-$plane (y\,=\,0\,mm) in the FoV calculated in MATLAB; the $\bm{B_z}$ along the 
  $x-$direction at y = 0, 20, 40, 60, 80\,mm (d) in the central $xy-$plane (z\,=\,0\,mm), and (e) in the side $xy-$plane (z\,=\,25\,mm) in the FoV.}
\label{fig:linear_one}
\end{figure}

If the requirement for homogeneity can be relaxed further with the implementation of ultra broadband techniques, a more linear field can be obtained from the GA optimization. Fig.\,\ref{fig:linear_one} shows the $\bm{B_0}$ field from a design with the relaxed field homogeneity but a much more uniform monotonic gradient pattern compared to the design presented in Section\, \ref{sec:results}. The preset $R^j_{in}$ ($j=1,2,\ldots,9$) were the same as those presented in Section\,\ref{sec:results}, and the optimized $R^i_{out}$ ($i=1,2,\ldots,12$) were [300 297.6 295.2 292.8 290.4 288.0 285.6 283.2 280.8 278.4 276.0 273.6] (unit:\,mm). This further optimized design provides a $\bm{B_0}$ field with an average field strength of 103.04\,mT, the field inhomogeneity of 178960\,ppm in FoV, and the gradient fields rang from $82.1$\,mT to $85.2$\,mT along the $x-$direction in the region -100\,mm$<x<$50\,mm in the central $xy-$plane (z\,=\,0\,mm) of the FoV. Compared to the results presented in Fig.\,\ref{field_plot_of_optimization_result}, with a compromised field strength and field homogeneity, the monotonic region is wider along the $x-$direction and the gradients are close to each other at different values of y. The uniformity of gradients along the $y-$direction helps to improve the uniformity of resolution in the reconstructed images\,\cite{cooley2018}. Fig.\,\ref{reconstructed_images}\,(d) shows the numerically reconstructed image by using the magnetic field in Fig.\,\ref{fig:linear_one}. In Fig.\,\ref{reconstructed_images}\,(d), clearer features are seen at the center of the image with the improved monotonicity compared to Fig.\,\ref{reconstructed_images}\,(b). The SSIM of the image using the further improved array is 0.453, which is an increase by 1.5\% and 41.1\% compared to that using the optimized array and the Halbach one in Fig.\,\ref{reconstructed_images}\,(b) and (c), respectively. However, it is worth pointing out that an improved monotonicity is obtained at a price of having a lower field strength and the lower field homogeneity, which may cause a low SNR and a consequent low spatial resolution, and a wide working frequency band for the RF system, respectively.  

\begin{figure}[t]
    \centering
        \includegraphics[height=0.22\linewidth]{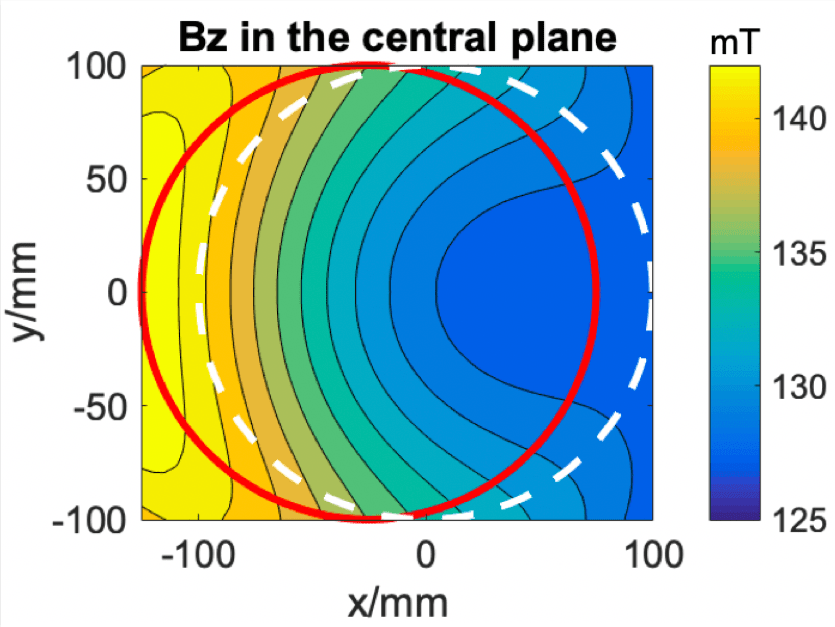}
  \caption{The red circle encloses the FoV with the center 25$\,$mm off the center of the array along the negative $x-$direction in the $xy-$plane (z\,=\,0\,mm), and the dashed white circle encloses the one without the offset distance.}
\label{illustration_of_offset}
\end{figure}
The FoV is 25$\,$mm off the center of the magnet array to have more regions with monotonic gradients, avoiding the area where the gradient is low (as shown in Fig.$\,$\ref{illustration_of_offset}). 
The magnet array will be rotated around the center of the FoV rather than the center of the magnet array to facilitate a non-Fourier transform image reconstruction. A rotation mechanism can be designed accordingly to achieve this off-center rotation.

\begin{figure}[t]
    \centering
    \subfloat[]{
        \includegraphics[height=0.2\linewidth]{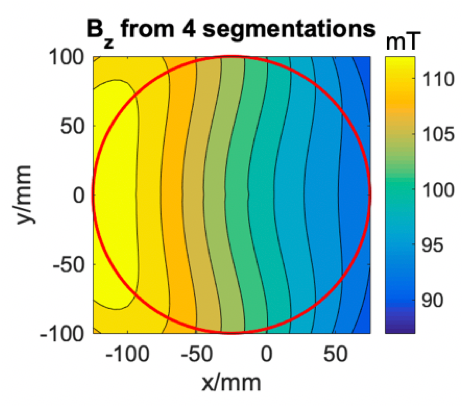}}
        \hspace{0.005\linewidth}
    \subfloat[]{
        \includegraphics[height=0.2\linewidth]{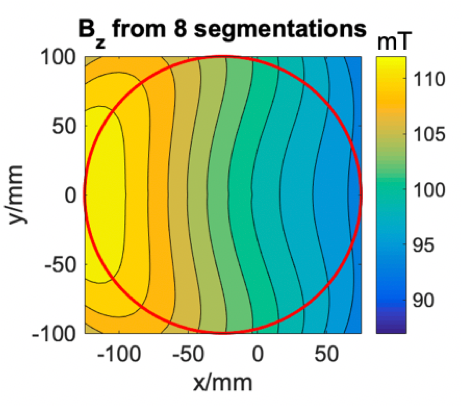}}
        \hspace{0.005\linewidth}
    \subfloat[]{
        \includegraphics[height=0.2\linewidth]{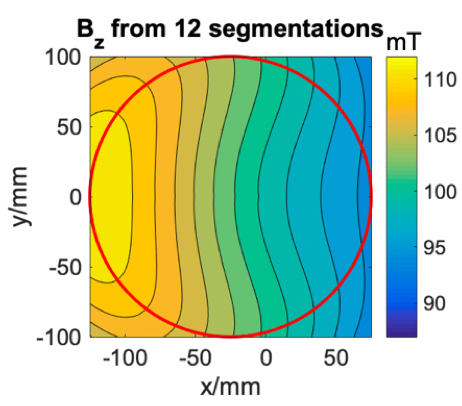}}
        \hspace{0.005\linewidth}
    \subfloat[]{
        \includegraphics[height=0.2\linewidth]{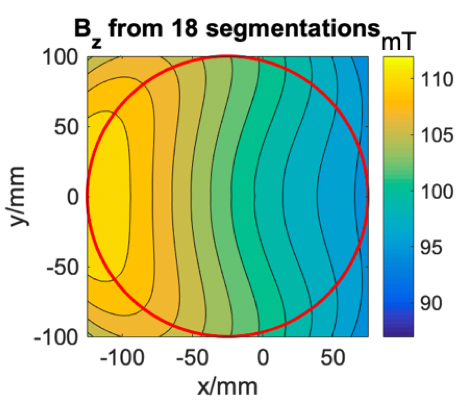}} 
        \\
    \subfloat[]{
        \includegraphics[height=0.2\linewidth]{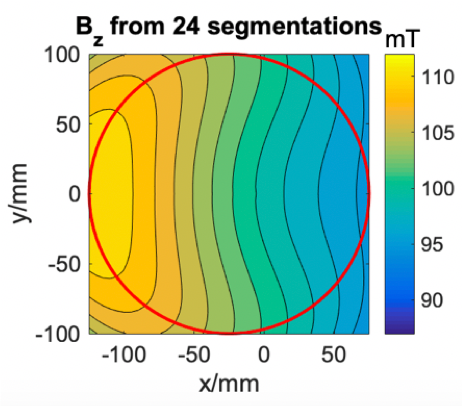}}
        \hspace{0.02\linewidth}
    \subfloat[]{
        \includegraphics[height=0.2\linewidth]{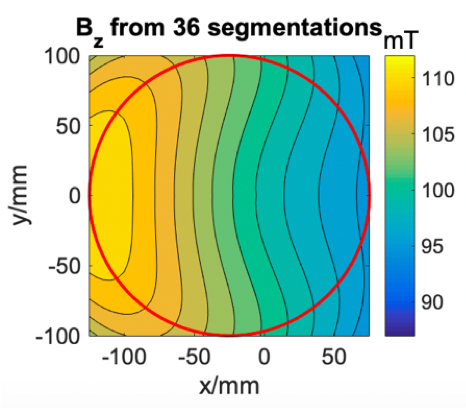}} 
        \hspace{0.02\linewidth}
    \subfloat[]{
        \includegraphics[height=0.2\linewidth]{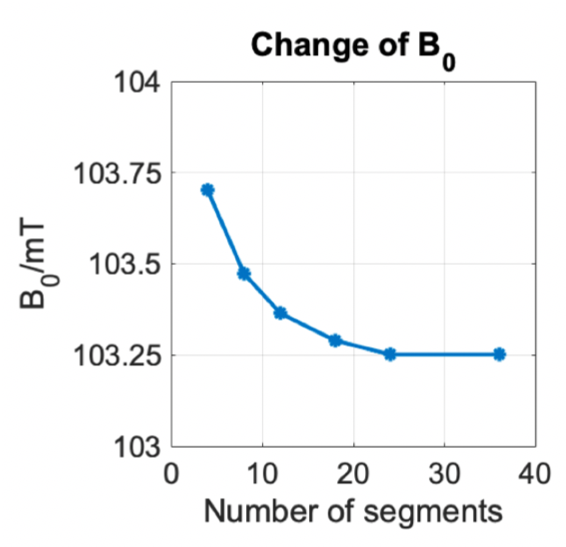}}
        \hspace{0.02\linewidth}
    \subfloat[]{
        \includegraphics[height=0.2\linewidth]{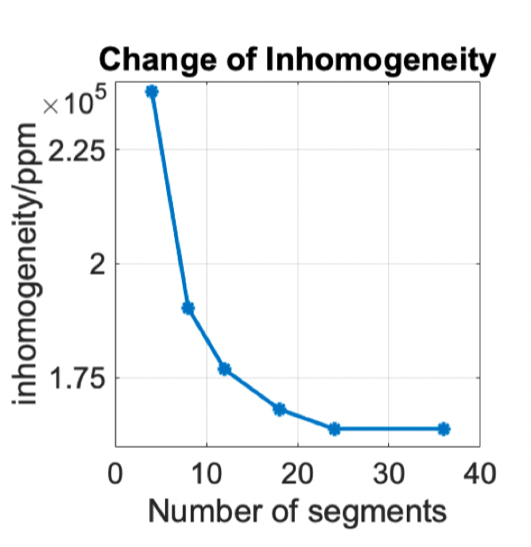}} 
  \caption{(a-f) The $\bm{B_0}$ field generated by the irregular-shaped ring-pair magnet array presented in Fig.\,\ref{fig:linear_one} with each annular magnet segmented to 4, 8, 12, 18, 24, and 36 fan-shaped magnets, respectively. (g-h) The changes of the average $\bm{B_0}$ field strength and the field inhomogeneity with the increase of number of segments.}
\label{fig:influence_of_segmentation}
\end{figure}
The less number of segments are helpful to reduce the cost of the whole magnet array. Here, the influence of the number of segments on the field pattern was investigated. Each annular magnet under optimization was segmented to 4, 8, 12, 18, 24, and 36 fan-shaped elements, and calculated using (\ref{eq4}) and (\ref{eq_BTotal}), respectively. In all six cases, the $R^i_{out}$ were tapered from 300\,mm to 273.6\,mm (as the design shown in Fig.\,\ref{fig:linear_one}\,(b) and (c)) according to the Eq.\,(\ref{eq:tapper_func}). The $\bm{B_0}$ field calculated in the central $xy-$plane (z\,=\,0\,mm) of FoV based on different segmentations are shown in Fig.\,\ref{fig:influence_of_segmentation}\,(a-f). The changes of the average field strength $\bm{B_{0-avg}}$ and field inhomogeneities (unit: ppm) with the increase of the number of segments were shown in Fig.\,\ref{fig:influence_of_segmentation}\,(g-h). As can be seen, the segmentation with different element numbers showed a similar field pattern and average field strength. However, the field inhomogeneity was influenced a lot by the number of segments as shown in Fig.\,\ref{fig:influence_of_segmentation}\,(h). Overall, $20\sim24$ will be a suitable range for the number of segments of an annular magnet in the proposed irregular-shaped ring-pair magnet array, which can provide both good field homogeneity and less segments to be calculated.

The fan-shaped magnets are widely used in the designs of motors and in those of accelerators. They are easy to fabricate. However, the optimized magnet array consists of 24 fan-shaped magnets of different dimensions, it would be costly for a physical implementation. The next step for this design is to explore the possibility of using magnet bars and cubes for a similar field pattern with a similar field strength and field homogeneity, probably through optimizing the locations of the magnets.

\section{Conclusion}\label{sec:conclusion}
We present a design and the optimization of an irregular-shaped ring-pair magnet array that generates a 1D monotonic field pattern for 2D head imaging. GA was applied for the optimization, and a current model for a fan-shaped ring segment pair was derived for a fast forward calculation for the optimization.In the simulation, the optimized magnet array shows an average $\bm{B}_0$ field of 132.98$\,$mT and the field homogeneity of 151840$\,$ppm in the a cylindrical FoV with a diameter of 200\,mm and a length of 50\,mm. Furthermore, the obtained magnetic field was applied as a SEM for signal encoding for imaging for a further evaluation of the proposed magnet array. Kaczmarz  iteration  method was used for the image reconstructions. The proposed magnet array successfully removes the blur at the central region of the image encoded by a Halbach array due to a highly monotomic gradient. It is a promising alternative to provide SEMs for imaging in a permanent-magnet-based low-field portable MRI system, besides a sparse Halbach array\,\cite{cooley2018} and an Aubert-ring-pair aggregate\cite{Ren2018}. 


\section*{Acknowledgements}

Z. H. Ren would like to thank the support from the Singapore University of Technology and Design President Graduate Fellowship. 
This work was supported by Singapore-MIT Alliance for Research and Technology Innovation Grant (ING137068).

\end{document}